%% file: main.tex
\documentclass{article}
\usepackage{spconf,amsmath,graphicx}
\usepackage{multirow, booktabs}
\usepackage{enumitem}
\usepackage{pifont}
\usepackage{xcolor}
\usepackage{balance}

\title{To Dereverb Or Not to Dereverb? \\
	Perceptual Studies On Real-Time Dereverberation Targets}
\name{Jean-Marc Valin, Ritwik Giri, Shrikant Venkataramani, Umut Isik, Arvindh Krishnaswamy}
\address{Amazon Web Services\\
Palo Alto, CA, USA\\
\small{\texttt{\{jmvalin, ritwikg, shriven, umutisik, arvindhk\}@amazon.com}}
}
\begin{document}
\ninept
\maketitle
\begin{abstract}
In real life, room effect, also known as room reverberation, and the present background noise degrade the quality of speech. Recently, deep learning-based speech enhancement approaches have shown a lot of promise and surpassed traditional denoising and dereverberation methods. It is also well established that these state-of-the-art denoising algorithms significantly improve the quality of speech as perceived by human listeners. But the role of dereverberation on subjective (perceived) speech quality, and whether the additional artifacts introduced by dereverberation cause more harm than good are still unclear. In this paper, we attempt to answer these questions by evaluating a state of the art speech enhancement system in a comprehensive subjective evaluation study for different choices of dereverberation targets.

\end{abstract}
\input{intro_v2}
\input{method_v2}
\input{targets}
\input{expr}
\input{conclusion}

\balance
\bibliographystyle{IEEEbib}
\bibliography{refs.bib}

\end{document}

%% file: intro_v2.tex
\section{Introduction}
\label{sec:intro}


Real-time denoising and dereverberation networks have started to attract a lot of attention from the speech and audio research community~\cite{xu2014regression, defossez2020real,isik2020poconet, wang2018supervised,giri2019attention}. In the case of speech denoising algorithms, these approaches have been demonstrated to significantly enhance the quality of the speech signals for human listeners. However, there has been relatively limited effort towards understanding the effects of room reverberation on the performance of these models and the resulting speech quality. In this work, we primarily focus on this aspect and analyze the effects of dereverberation on subjective speech quality.

Reverberation is the effect of the acoustic channel introduced into the speech signal as it travels from the source to the microphone or the hearing system. In a video-call setting, reverberation manifests in the form of reflections of the sound waves from the different surfaces of the room. This effect can be modeled as a linear time-invariant (LTI) system that is characterized by its impulse response (a.k.a, the room impulse response (RIR)). The initial samples of the RIR, known as \textit{early reflections}, capture the sparse reflections that arrive at the microphone within approximately 20~ms of the direct sound. They are followed by the more complex \textit{late reverberation}, which decays exponentially over hundreds of ms. It has been shown that late reverberation is primarily responsible for severely degrading speech quality and subsequently, speech intelligibility for human listeners~\cite{helfer1990hearing}.

Several methods based on classical signal processing have been proposed for speech dereverberation over the past decade~\cite{lebart2001new, wu2006two,  nakatani2010speech,yoshioka2012generalization}. For example,~\cite{wu2006two} is based on a two-stage approach, which cancels early reflections with an inverse filter and reduces late reverberations using spectral subtraction. Another line of work~\cite{nakatani2010speech,  yoshioka2012generalization} focuses on long-term linear prediction and has proven useful for late reverberation suppression. As is the case with speech denoising algorithms, deep-learning based supervised techniques have outperformed these classical approaches for speech dereverberation as well~\cite{han2015learning, kodrasi2018single, zhao2018late}. Mixtures containing speech and noise are convolved with synthesized RIRs to generate reverberant input mixtures that are fed into the network. The network is trained to produce the corresponding anechoic clean target speech as its output. Often these models are trained to estimate binary or ratio masks that can be superimposed on the Fourier representation of the noisy reverberant mixture to extract the Fourier representation of the anechoic target. Alternatively, in~\cite{williamson2017time, isik2020poconet}, the authors propose to perform dereverberation in the complex domain by estimating a complex ideal ratio mask.

When dealing with such reverberant environments, many deep-learning based dereverberation approaches attempt to perform a full dereverberation of the speech signal and retain only the direct path \cite{williamson2017time, ernst2018speech}. Since the full dereverberation task is significantly more difficult than denoising, these approaches often introduce undesired artifacts, hence degrading the quality of speech. This issue was further highlighted in the findings of the recently concluded Deep Noise Suppression Challenge~\cite{reddy2020interspeech}. Some participants~\cite{isik2020poconet,  defossez2020real} noted that partial dereverberation (by setting suitable training target) improved the speech quality in terms of  Mean Opinion Score (MOS), especially for highly reverberant cases. In other earlier works~\cite{zhao2018late,kodrasi2018single}, authors have proposed only suppressing the late reverberations to avoid any undesired artifacts.

In this paper, we address this issue in depth and analyze the effects of dereverberation on perceived speech quality. To do so, we train state-of-the-art dereverberation models with varying levels of reverberation on the training targets, ranging from no-dereverberation to full-dereverberation. We also experiment with different dereverberation strategies for these training targets. These models are compared on multiple tasks by performing a large-scale subjective evaluation study to identify the best dereverb strategy for optimal speech quality as perceived by human listeners. These listening tests also provide interesting insights into the relationship between room reverberation and the perceived naturalness of the speech signals. For example, our results reveal that partial dereverberation using an attenuated and decayed impulse response produces the highest MOS, especially in reverberant environment. To the best of our knowledge, we are not aware of any work that performs such a comparison of training targets for the dereverberation task.

For our experiments,  we use the recently proposed PercepNet architecture for real-time low-complexity speech enhancement~\cite{valin2020perceptually}. PercepNet is a perceptually motivated model that uses the spectral envelopes and the periodicity of the speech signals to perform state-of-the-art speech enhancement. It also exploits the power of a neural network by estimating the optimal gain function of a smoothed energy contour using a recurrent neural network (RNN).


The rest of the article is organized as follows: In Section
2, we briefly describe PercepNet and its components. In Section 3 we describe the different training targets used in our study. Section 4 presents the subjective evaluation results for different training targets, and finally, Section 5 concludes the paper and talks about some future research directions.

%% file: method_v2.tex
\section{System Overview}
\label{sec:sys}

Let $x(n)$ be a clean speech speech signal and $y(n)$ denote the corresponding noisy signal captured by a hands-free microphone in a noisy reverberant room. The noisy signal $y(n)$ can be expressed as
\begin{equation}
y(n) = x(n) \star h(n) + \eta(n)\ .
\end{equation}
Here, $\eta(n)$ denotes the additive background noise present in the room, $h(n)$ represents the impulse response from the talker to the microphone, and $\star$ denotes the convolution operator. The goal of the speech enhancement model is to estimate a function that will take some representation of the noisy signal $y(n)$ as its input and estimate the cleaned up version of the signal, i.e. denoised and dereverberated signal $\hat{x}= f(y)$ as its output.

\begin{figure}
	\centering{\includegraphics[width=1\columnwidth]{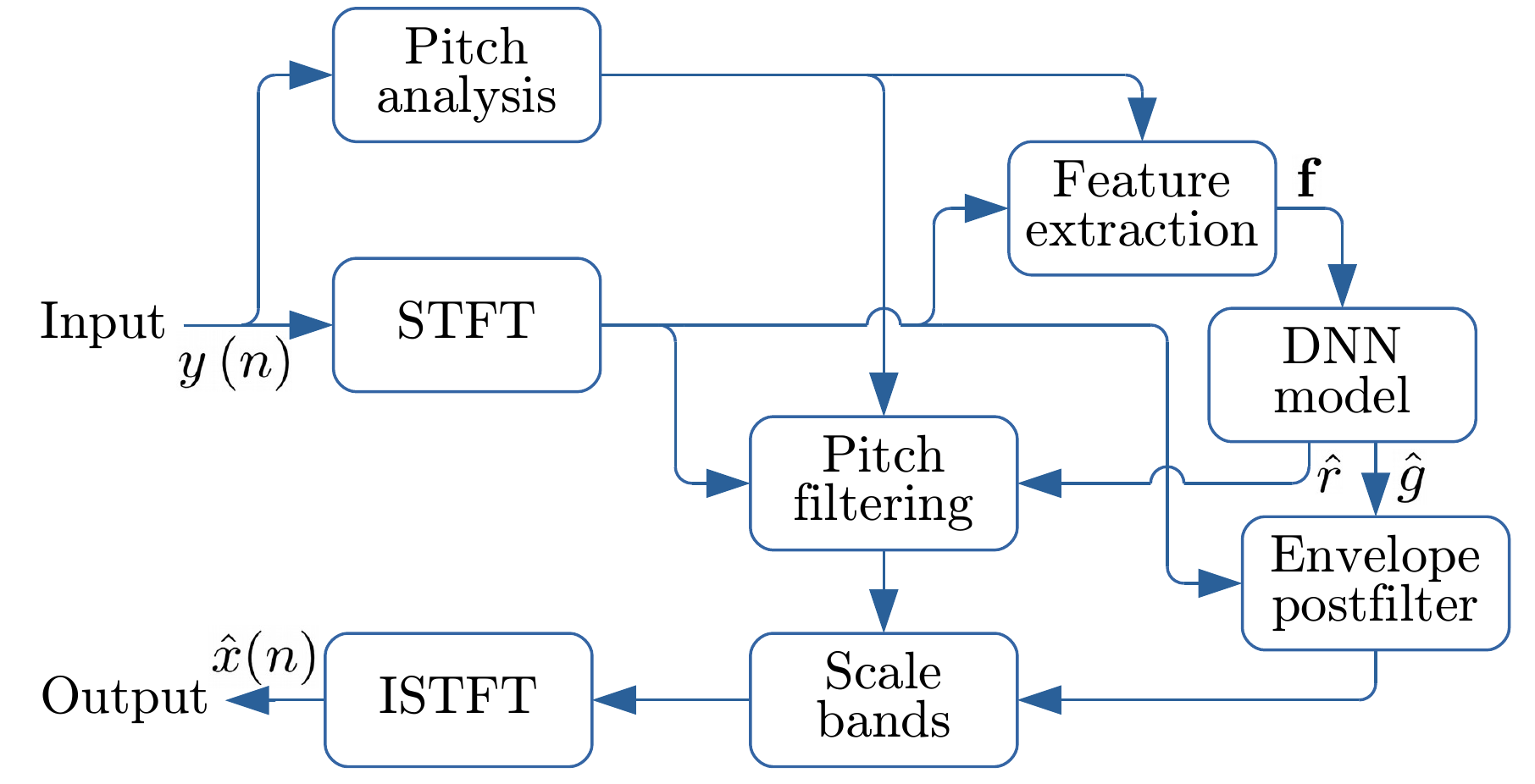}}\caption{Overview of the PercepNet algorithm.\label{fig:Overview-of-algorithm}}
	\label{percep:fig}
\end{figure}

In this work, we use the recently introduced real-time speech enhancement approach named PercepNet \cite{valin2020perceptually} as our speech enhancement model. We briefly describe the details of PercepNet in the following section.

\subsection{PercepNet: Background}
The PercepNet algorithm operates on 10-ms frames with 30 ms of look-ahead and enhances 48 kHz speech in real-time. Despite operating on a much lower complexity than the maximum allowed by the recently concluded first DNS challenge~\cite{reddy2020interspeech}, PercepNet ranked second in the real-time track. Fig.~\ref{percep:fig} presents an overview of the PercepNet algorithm. The two key elements of the algorithm are: (i) an RNN model to estimate band ratio masks, and (ii) a perceptually motivated pitch-filter.

Instead of operating on DFT bins (like most of the speech enhancement methods), PercepNet operates on only 32~triangular spectral bands, spaced according to the equivalent rectangular bandwidth (ERB) scale. PercepNet uses an RNN to estimate a ratio mask in each of these bands. This ratio mask can also be interpreted as the corresponding gain that needs to be applied to the noisy signal to match the spectral envelope of the clean target:
\begin{equation}
g_b(l) = \frac{X_b(l)}{Y_b(l)}\ ,
\end{equation}
where, $g_b(l)$ denotes the optimal gain in band $b$ for frame $l$. $X_b(l)$ denotes the the $L_2$ norm of the complex-valued spectrum of clean signal $x$ for band $b$ in frame $l$, and $Y_b(l)$ denotes the the $L_2$ norm of the complex-valued spectrum of the noisy signal $y$ for band $b$ in frame $l$.

To reconstruct the harmonic properties of the clean speech from the spectral envelopes, PercepNet also employs a comb filter controled by the pitch frequency. The use of such a time-domain comb filter allows us to achieve much finer frequency resolution than would otherwise be possible with the STFT (50~Hz using 20\nobreakdash-ms frames). The comb filter's effect is independently controled in each band using \textit{pitch filtering strength} parameters. Details of the comb filter can be found in~\cite{valin2020perceptually}.




%% file: targets.tex
\section{Training Targets}
\label{sec:expr}

One important question when attempting to dereverberate speech is
what we would like the dereverberated speech to sound like. Let $h_{0}\left(t\right)$
be the impulse response used to generate the input reverberant speech
$y\left(n\right)$ during training and $h_{1}\left(t\right)$ be the
impulse response used to generate the target speech $x\left(n\right)$.
Setting $h_{1}\left(t\right)=h_{0}\left(t\right)$ would lead to no
dereverberation. On the opposite end, setting $h_{1}\left(t\right)=\delta\left(t\right)$,
where $\delta\left(t\right)$ is the Dirac function, would lead to
complete dereverberation. This is not an ideal scenario for the following
reasons:
\begin{itemize}
\item Anechoic speech sounds highly unnatural;
\item Early reflections are much harder to remove, but have a lesser impact
on quality than late reverberation;
\item The difficulty of solving the problem leads to excessive artifacts
in the enhanced speech.
\end{itemize}
One possibility as proposed in~\cite{zhao2018late} is to keep the early reflections
and discard all late reverberation. In this work, we consider ways
of modifying the impulse response $h_{0}\left(t\right)$ to make the
task as easy as possible for the DNN, while also reducing the perceptual
impact of reverberation. Our goal is to shape $h_{0}\left(t\right)$
in the minimal way that leads to a target $h_{1}\left(t\right)$ that
has low perceived reverberation.

Considering $T_{0}=20\,\mathrm{ms}$ as the boundary between early
reflections and late reverberation, we first consider the decay function
\begin{equation}
D\left(t\right)=\left\{ \begin{array}{ll}
1\ , & t<T_{0}\\
10^{-3\left(t-T_{0}\right)/R_{D}}\ , & t>T_{0}
\end{array}\right.\ ,
\end{equation}
where $R_{D}$ is the time it takes the function to decay by 60~dB.

Let $R_{0}$ be the reverberation time ($\mathrm{RT_{60\,dB}}$) of
the room modeled by $h_{0}\left(t\right)$. Using the target $h_{1}\left(t\right)=D\left(t\right)h_{0}\left(t\right)$
is then equivalent to shrinking the room such that the reverbation time
$R_{1}$ of the target room satisfies
\begin{equation}
R_{1}^{-1}=R_{0}^{-1}+R_{D}^{-1}\ .
\end{equation}
For a highly-reverberant room, the target reverberation time will thus
never exceed $R_{D}$. For small rooms, $D\left(t\right)$ will also
have a smaller impact than for large rooms, which is a desired property. 

In addition to shrinking the target room with the exponential decay
function $D\left(t\right)$, we can \emph{move} the target microphone
closer to the speaker by attenuating the late reverberation part of
$h_{0}\left(t\right)$ by a factor $\alpha$. This can be done smoothly
between time $T_{0}$ and $T_{1}=30\,\mathrm{ms}$ using the attenuation
function
\begin{equation}
A\left(t\right)=\left\{ \begin{array}{ll}
1, & t<T_{0}\\
\frac{1+\alpha}{2}+\frac{1-\alpha}{2} \cdot \cos\frac{\pi\left(t-T_{0}\right)}{T_{1}-T_{0}},\  & T_{0}<t<T_{1}\\
\alpha, & t>T_{1}
\end{array}\right.\ .
\end{equation}

Under the free-field assumption, the sound intensity of the direct
path decays as $1/d^{2}$, so multiplying $h_{0}\left(t\right)$ by
$A\left(t\right)$ is equivalent to having the target distance $d_{1}=\alpha \cdot d_{0}$.
In practice, considering the early reflections, the intensity tends
to decay more slowly, so the effect on the target distance is greater. 

\begin{figure}
       \centering{\includegraphics[height=18em,width=0.95\columnwidth]{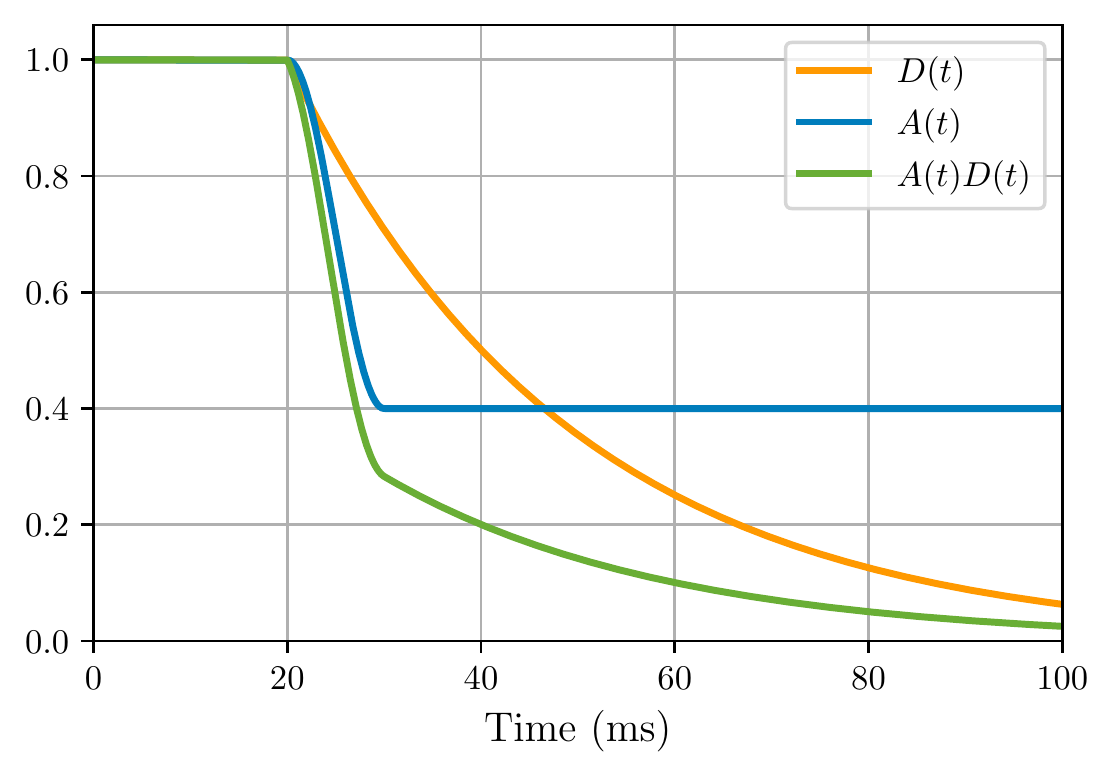}}\caption{RIR shaping functions.}
       \label{fig:target_scaling}
\end{figure}

The effect of the shaping functions described above is shown in Fig.\ \ref{fig:target_scaling}. We introduce and study four different training targets for dereverberation. The choice of the training target determines the ground truth gains $g_b(l)$, shown in Equation (2), which are used to train the DNN model.

\subsection{Full Dereverberation}

In a first experiment, we aim to remove all late reverberation, and only keep the direct path and early reflections in the estimated output. This is done by using
\begin{equation}
h_1(t) = A(t) h_0(t)\ ,
\end{equation}
with $\alpha=0$. We have two conditions for this experiment: $T_0=20\ \mathrm{ms},\ T_1=10\ \mathrm{ms}$, as well as $T_0=10\ \mathrm{ms},\ T_1=5\ \mathrm{ms}$. 

\subsection{Decayed Dereverberation}

For the full dereverberation model, the lack of any reverberation can make the enhanced signal sound overly dry and unnatural. To address this issue, we use the decay
\begin{equation}
h_1(t) = D(t)h_0(t)\ ,
\end{equation}
with $R_D=200\ \mathrm{ms}$. This shrinks the target room to a maximum of 200~ms reverberation time.

\subsection{Attenuated and Decayed Dereverberation}

While the partial dereverberation approach addresses the excessive dryness of the target, it may leave too-much reverberation in the output. Hence, in this case, we use the late reflection attenuation to bring the target microphone closer to the speaker using
\begin{equation}
h_1(t) = A(t) D(t) h_0(t)\ ,
\end{equation}
with $\alpha=0.4$ (-8~dB) and $R_D=200\ \mathrm{ms}$. For a speaker standing 2~m from the microphone, the target reverberation will sound as if the microphone were at most 0.8~m away (likely less in practice).

\subsection{No Dereverberation}

The "no dereverberation" case corresponds to $h_1(t) = h_0(t)$, so the target is a clean reverberant signal without any background noise.

\begin{figure}
       \centering{\includegraphics[height=18em,width=0.95\columnwidth]{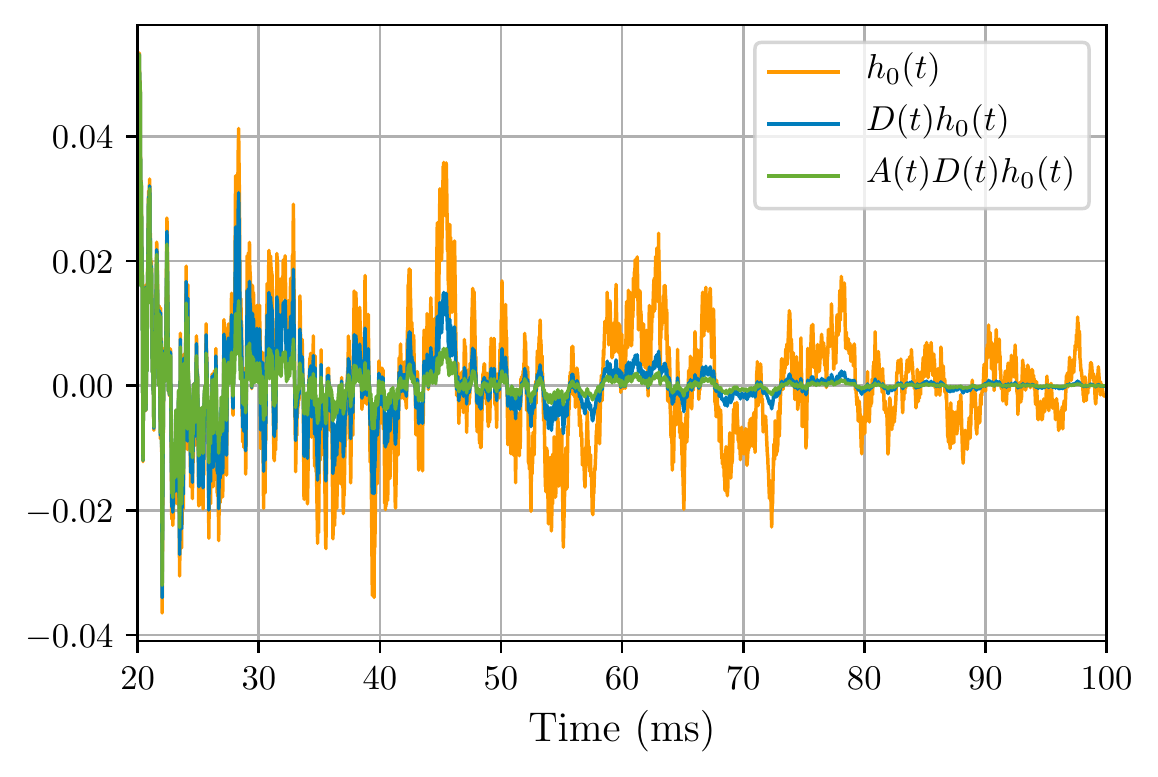}}\caption{RIR for clean speech label for three different training targets.}
       \label{target:fig}
\end{figure}

In Fig.~\ref{target:fig}, we present the late reflections component of the target label's RIR for three different training targets. Note that we did not include the early reflections component, i.e., the first 20~ms of RIR since that is unchanged for all 4 cases. 

%% file: expr.tex
\begin{table*}[hbt!]
	\caption{P.808 MOS results on DNS ICASSP 2021 blind test set. The 95\% confidence interval is 0.03.}
	\vspace{1em}
	\centering{
	\begin{tabular}{lcccccc}
		\hline \\[-2ex]
		$\,$ & Singing & Tonal & Non-english & English & Emotional & Overall\\
		\hline \\[-2ex]

		Noisy  & 2.96 & 3.00 & 2.96 & 2.80 & 2.67 & 2.86\\
		DNS Challenge Baseline \cite{braun2020data}  & 3.10 & 3.25 & 3.28 & 3.30 & 2.88 & 3.21 \\
		\textbf{Attenuated+decayed dereverb}  & \textbf{3.16} & \textbf{3.40} & \textbf{3.42} & \textbf{3.47} & \textbf{2.90} & \textbf{3.34} \\
		\hline
	\end{tabular}}
	\vspace{0.05in}
	\label{table:mos}
\end{table*}

\begin{table}
\caption{P.808 MOS results of different algorithms on the DNS INTERSPEECH 2020 Challenge synthetic development set (450 samples) and the DNS ICASSP 2021 blind test set (700 samples). The 95\% confidence interval is 0.04.}
	\vspace{1em}
	\centering{%
		\begin{tabular}{lcc}
			\hline
			Algorithm  & \textbf{MOS (P.808)}\tabularnewline
			\hline
			Noisy input & 2.95\tabularnewline
			DNS Challenge Baseline~\cite{braun2020data} & 3.21\tabularnewline
			No dereverb & 3.31\tabularnewline
			Full dereverb (20\ ms) & 3.35\tabularnewline
			Full dereverb (10\ ms) & 3.33\tabularnewline
			Decayed dereverb & 3.36\tabularnewline
			\textbf{Attenuated+decayed dereverb} & \textbf{3.36}\tabularnewline
			\hline
	\end{tabular}}
	\label{table:dev}
\end{table}

\begin{table}
\caption{P.808 MOS results of different reverberation targets reverberant samples. The 95\% confidence interval is 0.04.}
	\vspace{1em}
	\centering{%
		\begin{tabular}{lcc}
			\hline
			Algorithm  & \textbf{MOS (P.808)}\tabularnewline
			\hline
			No dereverb & 3.05\tabularnewline
			Full dereverb (20\ ms) & 3.38\tabularnewline
			Full dereverb (10\ ms) & 3.37\tabularnewline
			Decayed dereverb & 3.37\tabularnewline
			\textbf{Attenuated+decayed dereverb} & \textbf{3.39}\tabularnewline
			\hline
	\end{tabular}}
	\label{table:reverb}
\end{table}

\section{Experiments And Results}

We train on approximately 120~hours of clean speech recordings comprising over 200~speakers in more than 20~different languages. We also include the emotional speech training examples provided by DNS challenge organizers and the vocal tracks from ccMixter dataset~\cite{liutkus2014kernel} to our clean speech data. For the noise examples, we use a dataset that consists 80~hours of ambient noise recordings of various types. The speech and noise examples were sampled at 48~kHz. To simulate the effect of a reverberant environment, we use both real-recorded and synthetic room impulse responses (RIRs). We generate mixtures of clean reverberant speech convolved with an RIR and noise such that the SNRs range from -5~dB to 45~dB. We also randomly introduce a few noise-free examples during training. 

Even though multiple objective metrics exist to measure the quality of speech, it has been shown in~\cite{reddy2019scalable} that these objective metrics are not very reliable, and they correlate poorly with subjective MOS. Hence, in this work, we compare different training targets using a large scale MOS study. We use the methodology of ITU-T P.808 Subjective Evaluation of Speech Quality with a crowdsourcing approach~\cite{P.808}.

To demonstrate the effect of different dereverb training targets, we use the synthetic development eval set provided during the INTERSPEECH 2020 DNS Challenge~\cite{reddy2020interspeech}. Specifically, we evaluate our models on three different subsets: 
\begin{enumerate}
    \item Synthetic recordings of speech in background noise without room reverberation
    \item Synthetic speech recordings in a reverberant environment without any ambient noise, and
    \item Synthetic reverberant recordings in the presence of ambient noise.
\end{enumerate}
Each of the subsets comprises 150~utterances. The MOS scores are aggregated based on ten listening tests for each model output corresponding to each of these $150 \times 3 = 450$ synthetic inputs. We also additionally evaluate different training targets on the ICASSP 2021 DNS Challenge blind test set comprising 700~utterances. We wish to specifically emphasize that the evaluation of the models on the ICASSP 2021 DNS blind test set was performed after we submitted our processed outputs to the challenge. The model that was submitted to the challenge operates in real-time using only $4.5\%$ of an  i7-8565U core at 1.80GHz. 

Table~\ref{table:mos} shows the subjective MOS results measured by the ICASSP 2021 DNS challenge organizers over their blind test set (700~samples). Based on internal testing (without looking at the blind test set), we chose to submit the "Attenuated+decayed dereverb". The results clearly show that the proposed approach outperforms the challenge baseline in all categories.

Table~\ref{table:dev} includes results on the previous INTERSPEECH 2020 challenge development set for the different training targets. These samples include both reverberant and non-reverberant samples. All tested targets significantly exceed the quality of both the NSnet2 baseline and the noisy speech. Although the proposed partial dereverberation target produced the highest score, the difference between the evaluated targets are not statistically significant ($p>0.05$). This is in part due to the fact that many of the samples have little reverberation.

For that reason, we conducted an additional listening test comparing the five dereverberation targets on 600~reverberant samples. These include 300~synthetic reverberant samples from the INTERSPEECH 2020 DNS Challenge~\cite{reddy2020interspeech}, as well as 300~randomly-selected reverberant samples from the multi-channel Wall Street Journal Audio-Visual (MC-WSJ-AV) database. The results of the second test in Table~\ref{table:reverb} shows that dereverberation significantly improves the quality of reverberant speech for all targets tested. The second experiment again shows the "Attenuated+decayed dereverb" target is producing the best results. Although these results are again not statistically significant, they still provide an indication that it is a promising option. 

%% file: conclusion.tex
\section{Conclusion}
\label{sec:conc}

For supervised speech enhancement, the choice of an appropriate training target is critical. In this paper, we present and compare  different training targets for supervised dereverberation, and perform a large scale subjective evaluation of our choices. Our results indicate that training with a decayed and attenuated dereverberation target produces the highest quality, but more experiments would be needed to confirm that. 
